\documentclass[10pt,conference,final,a4paper]{IEEEtran}
\renewcommand{\baselinestretch}{1.07}

\usepackage{amsmath,graphicx,bm}
\usepackage{amssymb}
\usepackage{epsfig}
\usepackage[noadjust]{cite}
  
\ifCLASSINFOpdf
\else
\fi

\begin{document}
\title{Linear and Circular Microphone Array for Remote Surveillance: Simulated Performance Analysis}

\author{\IEEEauthorblockN{Abdulla AlShehhi, M. Luai Hammadih, M. Sami Zitouni, Saif AlKindi, Nazar Ali, and Luis Weruaga}
\IEEEauthorblockA{Department of Electrical and Computer Engineering \\
Khalifa University of Science, Technology and Research \\
Sharjah, United Arab Emirates}}


\maketitle

\begin{abstract}
Acoustic beamforming with a microphone array represents an adequate technology for remote acoustic surveillance, as the system has no mechanical parts and it has moderate size. However, in order to accomplish real implementation, several challenges need to be addressed, such as  array geometry, microphone characteristics, and the digital beamforming algorithms. This paper presents a simulated analysis on the effect of the array geometry in the beamforming response. Two geometries are considered, namely, the linear and the circular geometry. The analysis is performed with computer simulations to mimic reality. The future steps comprise the construction of the physical microphone array, and the software implementation on a multi-channel digital signal processing (DSP) system.
\end{abstract}

% \fi
\IEEEpeerreviewmaketitle

\section{Introduction}

Array beamforming \cite{Johnson1993} is a promising technique for audio surveillance and multi-conferencing. This technique enables the capture of a desired sound arriving from a certain spatial direction while suppressing undesired sounds coming from other directions . Applications such as acoustic surveillance, i.e. the unobtrusive monitoring and acoustic tracking of people located at relatively far distances and in actual noisy environments, can benefit from sound beamforming systems as these microphone arrays do not have mechanical movable parts. Certain compact array geometries make it possible for such microphone array systems to be implemented and used effectively in acoustic surveillance and multi-conferencing.
\\ \indent
Acoustic beamforming has been thoroughly studied \cite{Bradstein2001},\cite{Vary2006}. A microphone array is composed of several microphones aligned in a specific pattern or geometry to operate as a single device. Microphone arrays appeared nearly three decades ago, but this is still a field with active research \cite{Huang2011,Benesty2008,Tamai2003,Pessentheiner2012,Chen2009,Mabande2009,Liu2010}. A microphone array has the capability of separating between various signals by utilizing some techniques of filtering, substantially boosting the signal to noise ratio (SNR). The reading of all microphones are sampled and processed on a digital signal processing (DSP) computer to obtain the relevant signal and discarded the unwanted one. It can be utilized for various applications, for instance, finding the location of sound sources or distinguish between sounds according to direction without the need to physically move the array of microphones.
\\ \indent
The aim of this paper is to explore and understand the two main microphone array geometries, namely linear and circular, implement then in a software simulation environment, and carry out a quantitative and qualitative comparative performance analysis. The paper is structured as follows: Firstly, an illustration of the principles of beamforming is presented based on two structures, linear array beamformer and circular array beamformer in order to compare between them. After that, results of simulation done based on linear array and circular array beam-pattern are provided to verify which configuration gives the best performance. Finally the further work and conclusions close the paper.

%\psdraft
\begin{figure}[!b]
\centering
\includegraphics[width=75mm]{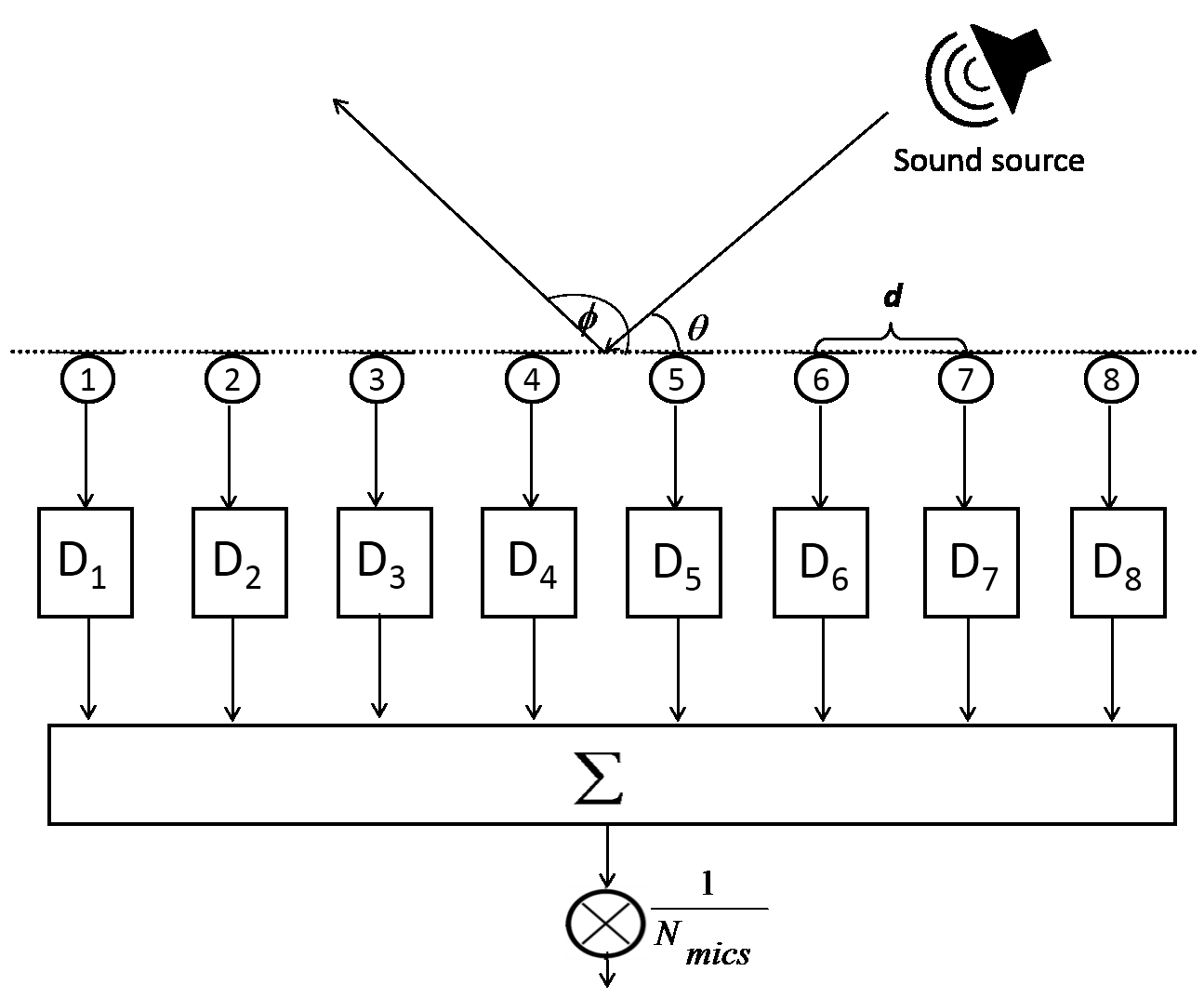}
\caption{Uniform linear $8$-microphone array. Angle $\phi$ corresponds to the steering angle and $\theta$ to the direction of sound arrival.}
\label{fig_linarray}
\end{figure}

\section{Principles of Beamforming}

\subsection{Linear Array Beamformer}

Figure \ref{fig_linarray} illustrates schematically the basic delay-and-sum (DaS) beamformer on a linear array.  The sound source given by $s(t)$ arrives at the direction given by $\theta$, while the microphone array has been set up to point to the direction $\phi$. In far-field conditions (the distance of the sound source is much larger than the array size), the signal captured by by $k$th microphone results in
\begin{equation}
x_k (t)= s(t-\Delta_k )
\end{equation}
\noindent where $\Delta_k$ represents the physical time delay at the $k$th microphone. It is simple to deduce that for such a geometric arrangement, the delay $\Delta_k$ corresponds to 
\begin{equation}\label{physdelay1}
\Delta_k = (d/c)\bigl(k- (N+1)/2\bigr)\cos\theta
\end{equation}
\noindent where $d$ is the distance between adjacent microphones, $c$ is the speed of sound in air, $N$ is the number of microphones, and the microphone index $k$ runs as $k = 1,\cdots,N$. Note that the signal at the microphone is then sampled and sent to a digital signal processor. However, for the sake of simplicity in the exposition, we will use continuous time in the subsequent analysis. The $k$th channel is then digitally delayed as
\begin{equation}
z_k (t)= x_k(t+\delta_k)
\end{equation}
\noindent where $\delta_k$ corresponds to the digital delay applied to the $k$th channel. Since the array is meant to be pointing to angle $\phi$, the digital delay $\delta_k$ can be easily obtained from \eqref{physdelay1}, that is,
\begin{equation}\label{digdelay1}
\delta_k = (d/c)\bigl(k- (N+1)/2\bigr)\cos\phi.
\end{equation}
\noindent Finally, all channels are averaged, this resulting in the final DaS signal as follows
\begin{equation}\label{average}
y(t) = \frac{1}{N}\sum_{k=1}^{N}{z_k (t)} = \frac{1}{N} \sum_{k=1}^{N}{s(t-\Delta_k+\delta_k)}.
\end{equation}
\noindent If the steering angle $\phi$ is set to the actual direction of the incoming sound $\theta$, the physical and digital delays compensate ($\delta_k = \Delta_k$), and the beamformer output is equal to the sound source signal, $y(t)=s(t)$. In general, the averaging among the signals at the adder results in the frequency-selective attenuation of the signal, as the translation of the beamformer output \eqref{average} to the frequency domain reveals
\begin{equation}
Y(\omega) = H_\phi(\omega,\theta) S(\omega).
\end{equation}
\noindent Here $\omega$ is frequency, $Y(\omega)$ and $S(\omega)$ correspond to the Fourier transform of $y(t)$ and $s(t)$ respectively, and $H_\phi(\omega,\theta)$ is the transfer function of the beamformer with steering angle $\phi$
\begin{equation}\label{filterbeam}
H_\phi(\omega,\theta) = \frac{1}{N} \sum_{k=1}^{N}{\exp\bigl(-j\omega(\Delta_k-\delta_k)\bigr)}.
\end{equation}
 \indent %
It is immediate to see that if the incoming sound arrives at the steering angle $\theta = \phi$, then $\Delta_k=\delta_k$, hence $H_\phi(\omega,\theta)=1$, that is, the sound signal is captured without spectral distortion. The linear geometry presents unfortunately an additional situation at which $\Delta_k=\delta_k$, which is for $\theta = -\phi$. This ``ambiguity" lobe is as dominant as the main one. In any other case $\theta \neq \phi$, the beamformer output will undergo a certain frequency-selective attenuation according to \eqref{filterbeam}.

\subsection{Circular Array Beamformer}

\begin{figure}[!t]
\centering
\includegraphics[width=89mm]{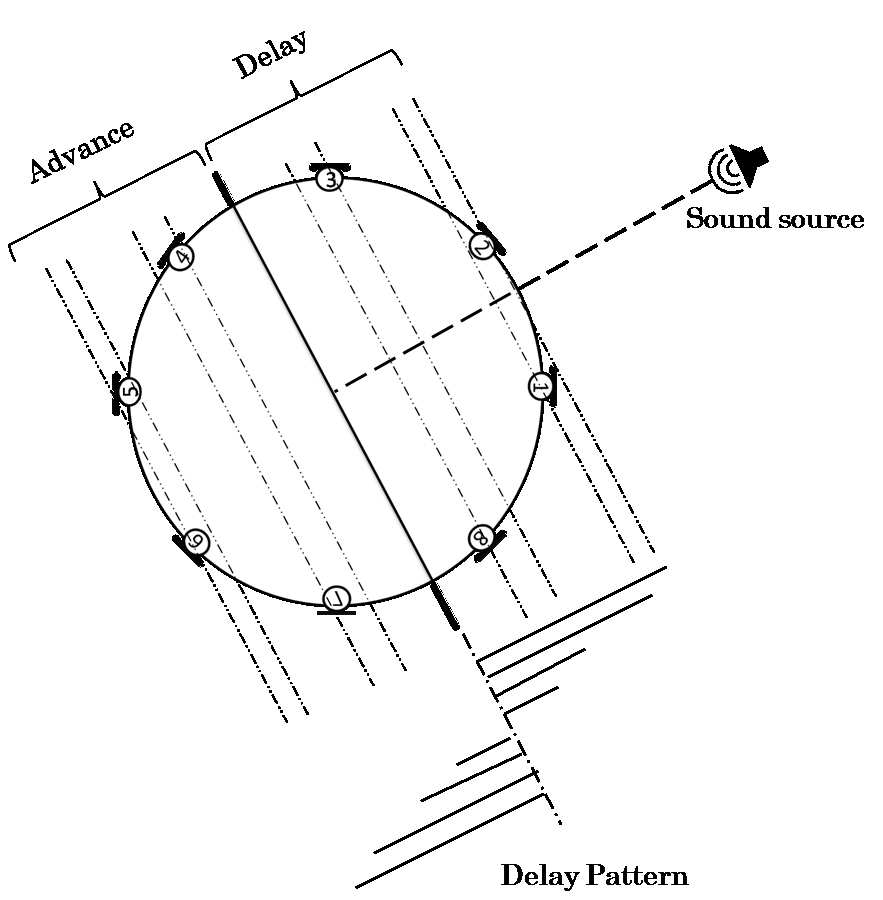}
\caption{Uniform circular $8$-microphone array.}
\label{fig_circarray}
\end{figure}

Figure \ref{fig_circarray} depicts the same DaS principle in a microphone array with circular geometry. This figure illustrates also the delay pattern at all the microphones: the delay pattern follows a trigonometric relation determined by the angle of each microphone $k2\pi/N$ with respect to the direction of arrival $\theta$. The physical delay that the incoming sound experiences at each microphone depends on the difference between the previous angles, that is,
\begin{equation}\label{physdelay2}
\Delta_k = (r/c)\cos\bigl(\theta-k2\pi/N\bigr)
\end{equation}
\noindent where $r$ is the radius of the circular geometry. It is important to relate the radius $r$ with the distance between adjacent microphones $d$, a relation that follows
\begin{equation}
d = 2 r \sin(\pi/N)
\end{equation}
\indent %
The digital delays follow the same trigonometric rule \eqref{physdelay2} depending on the steering angle $\phi$
\begin{equation}\label{digdelay2}
\delta_k = (r/c)\cos\bigl(\phi-k2\pi/N\bigr).
\end{equation}
\noindent Unlike in the linear case, the circular geometry does not present any ``ambiguity" lobe: the equation $\delta_k = \Delta_k$, that is,
\begin{equation}
\cos\bigl(\phi-k2\pi/N\bigr) = \cos\bigl(\theta-k2\pi/N\bigr)
\end{equation}
\noindent has only one solution, namely, $\theta=\phi$.

%\psdraft

\section{simulation Results}

This part sheds light on the differences between a uniform linear and circular $8$-microphone array by means of the respective simulated beam pattern and beam patterns in polar coordinates at three different single frequencies.

%
%\psdraft
\begin{figure}[!t]
\centering
\includegraphics[width=88mm]{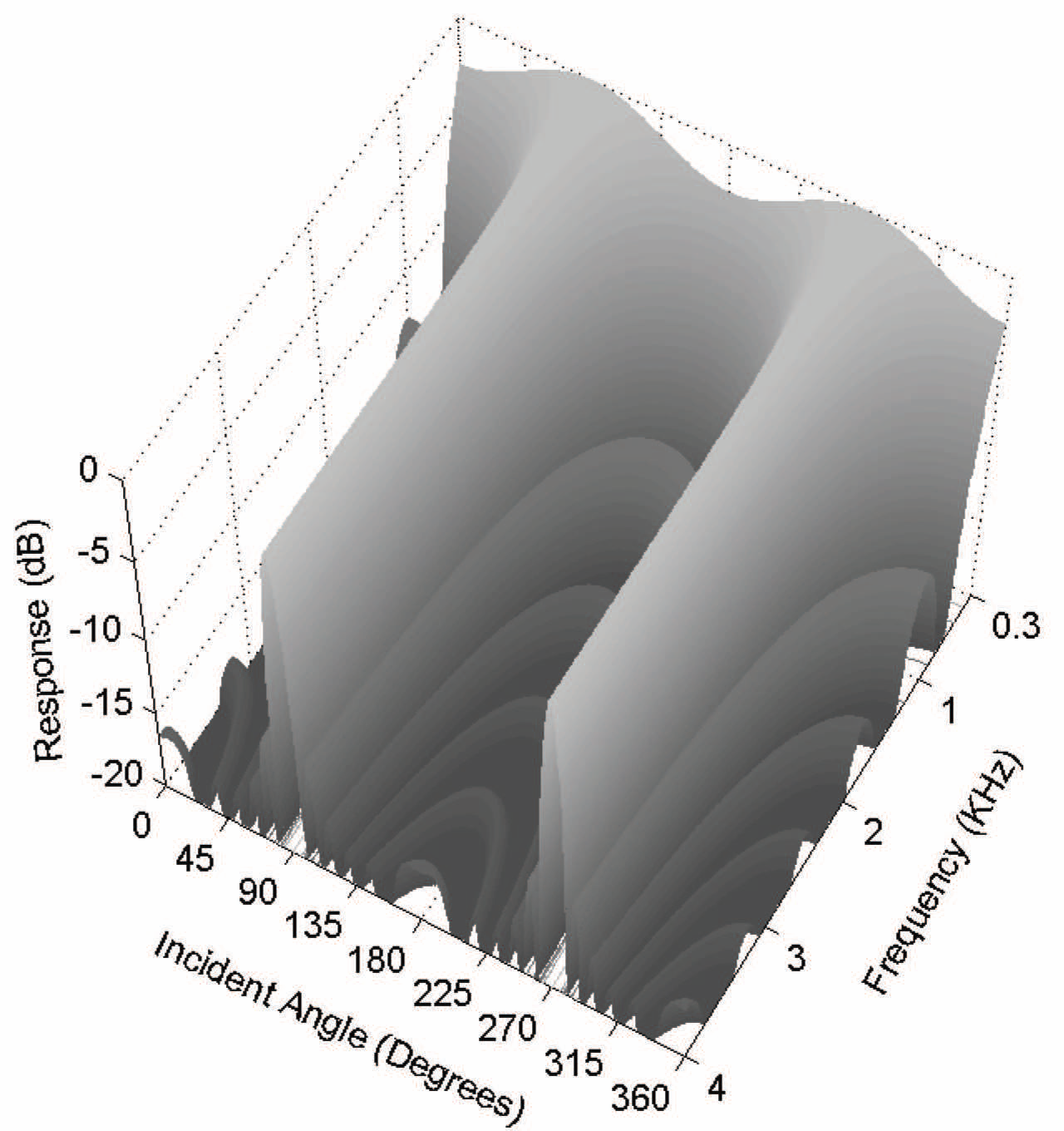} 
\caption{Beam pattern of a linear $8$-microphone array}
\label{fig_linearBP}
\end{figure}

%\psdraft
\begin{figure}[!t]
\centering
\includegraphics[width=88mm]{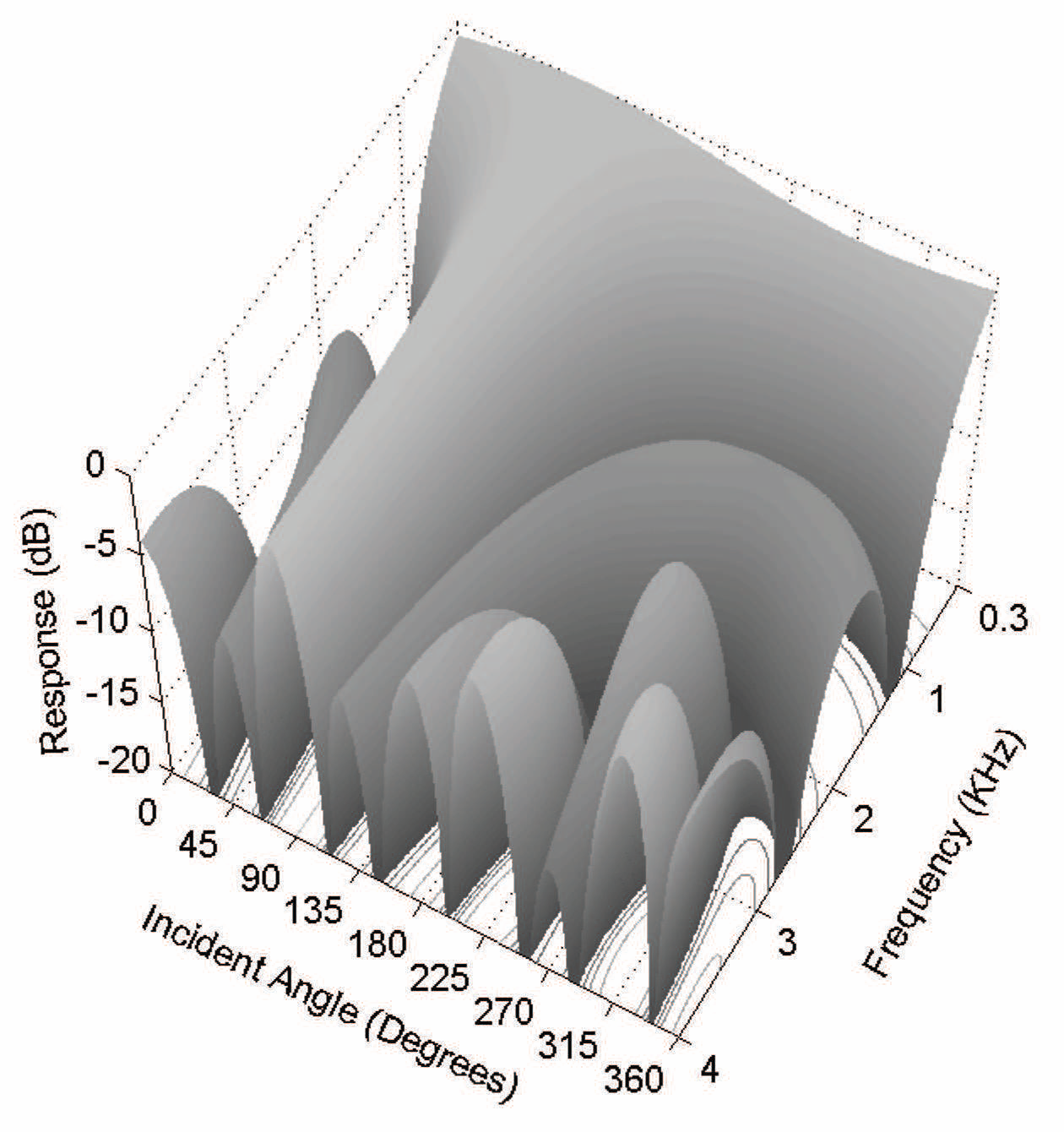} 
\caption{Beam pattern of a circular $8$-microphone array.}
\label{fig_circularBP}
\end{figure}

%\psdraft
\begin{figure}[!t]
%\centering
\raggedleft
\hspace{-6mm}
\begin{minipage}{94mm}
\includegraphics[width=47mm]{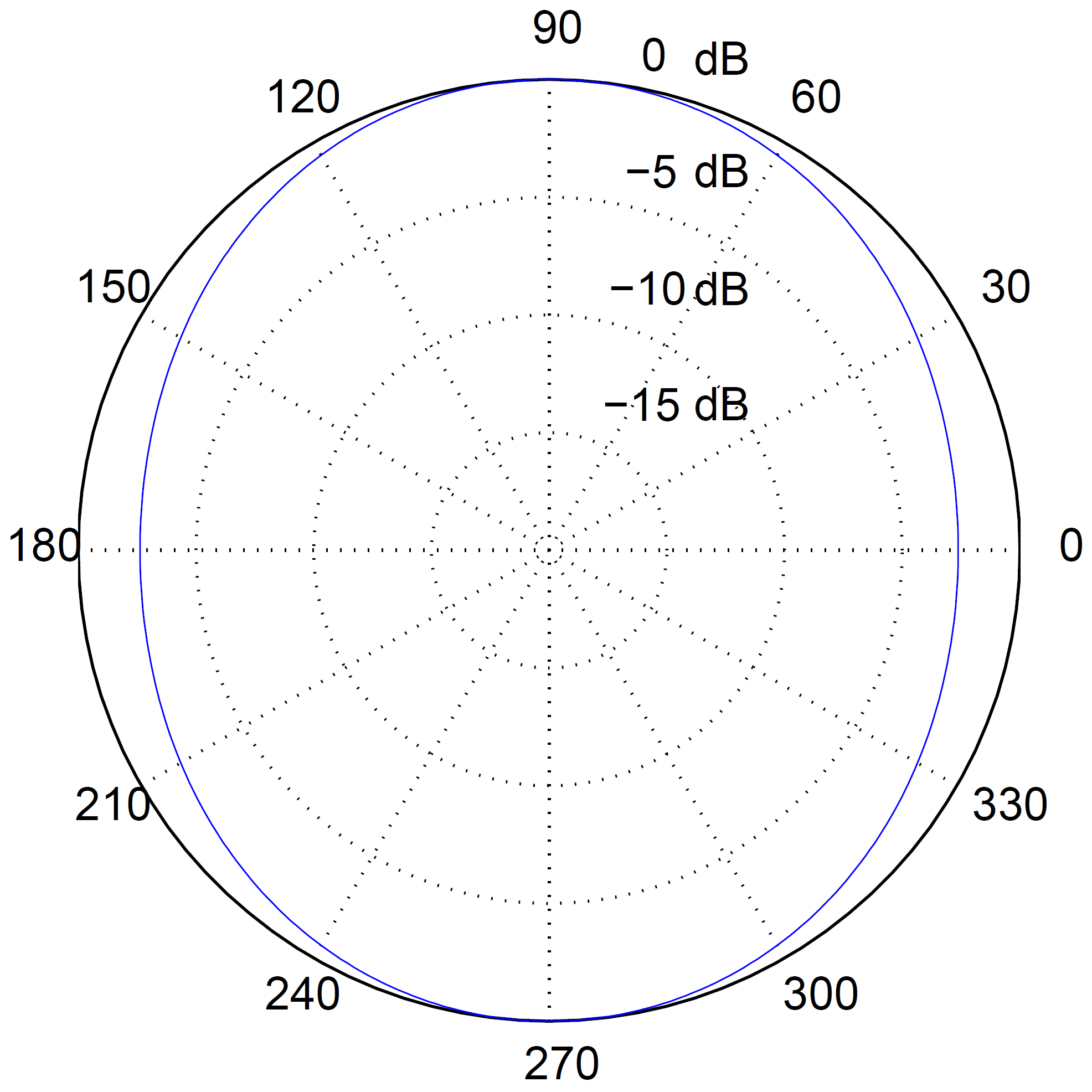}\includegraphics[width=47mm]{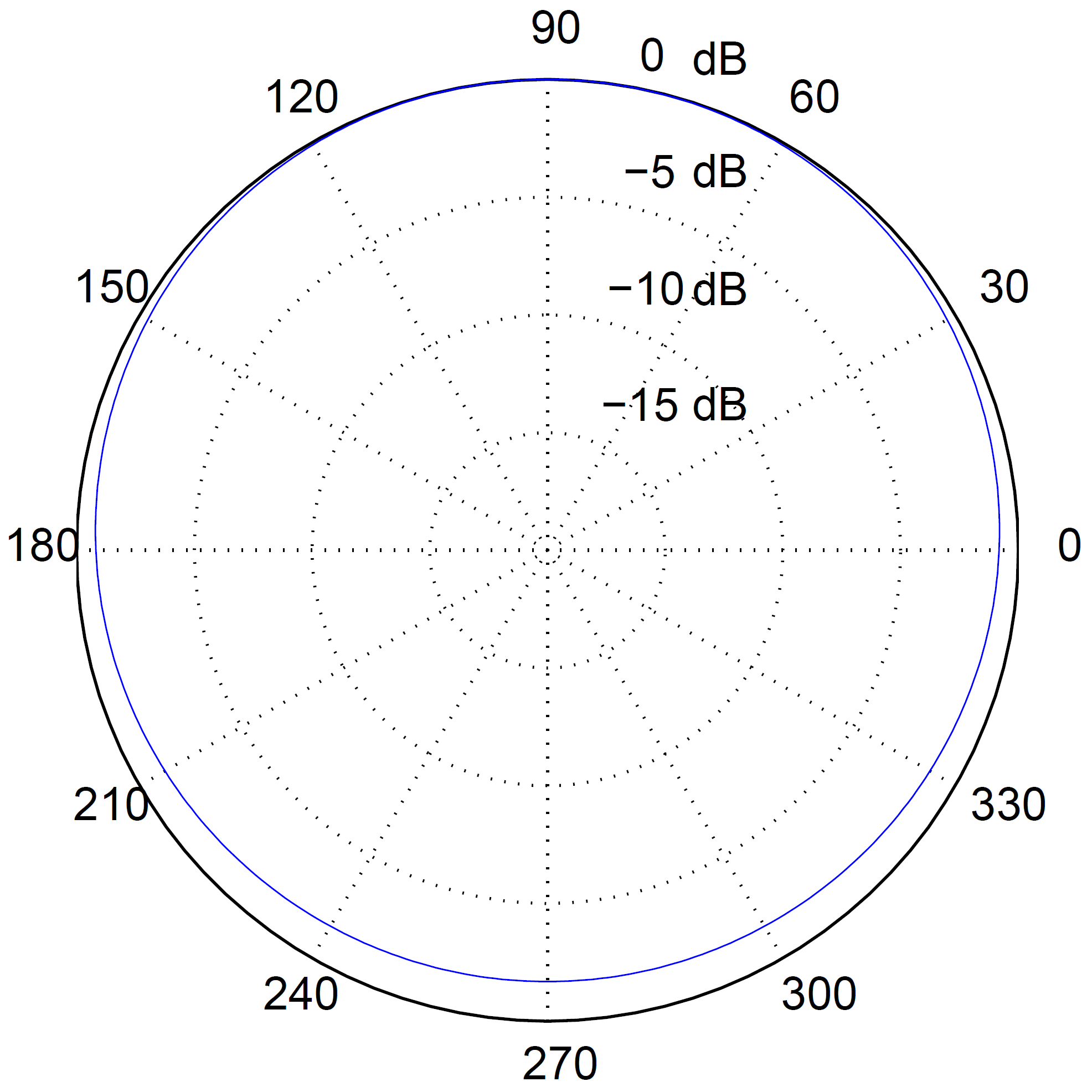}\\
\includegraphics[width=47mm]{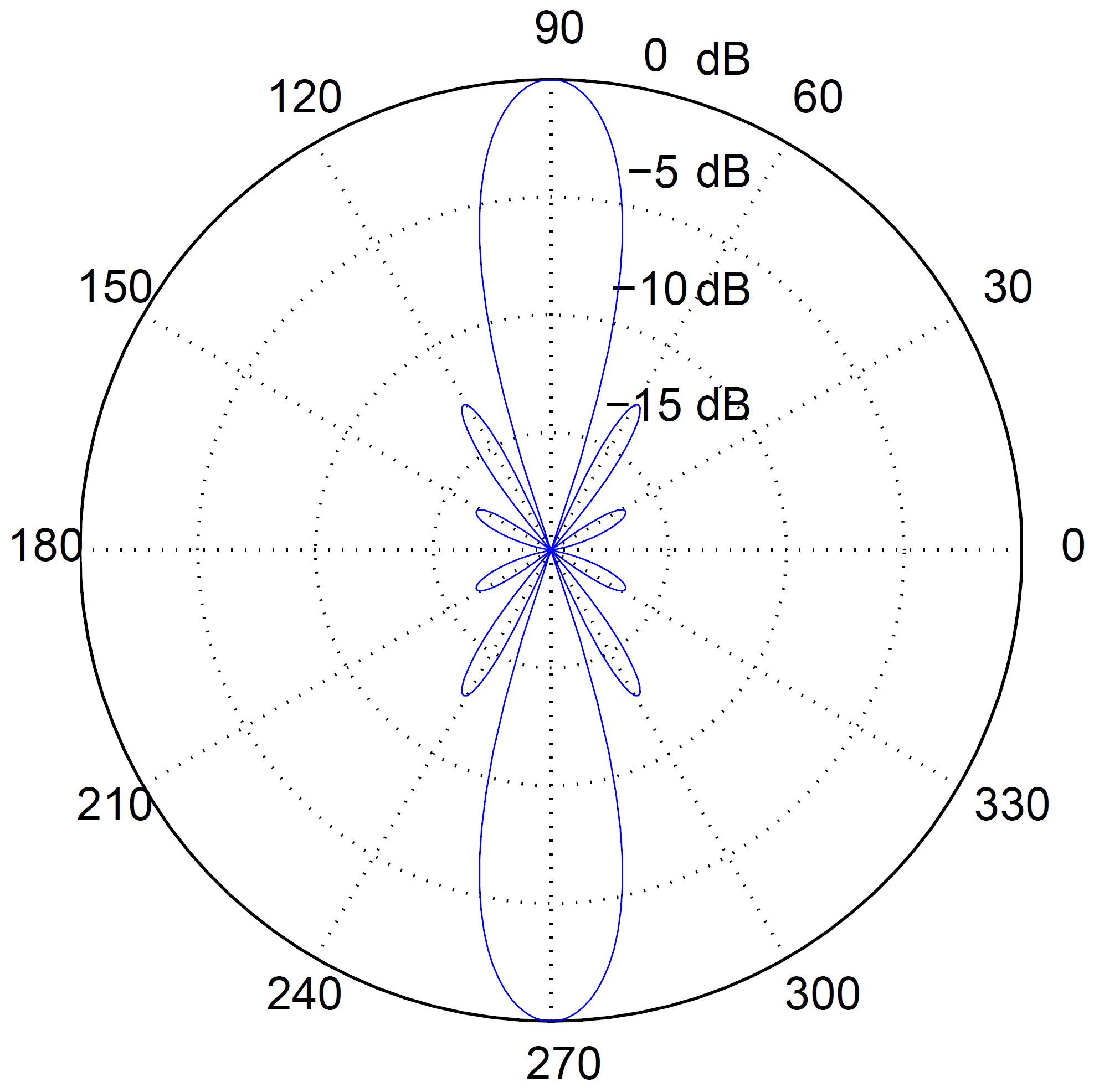}\includegraphics[width=47mm]{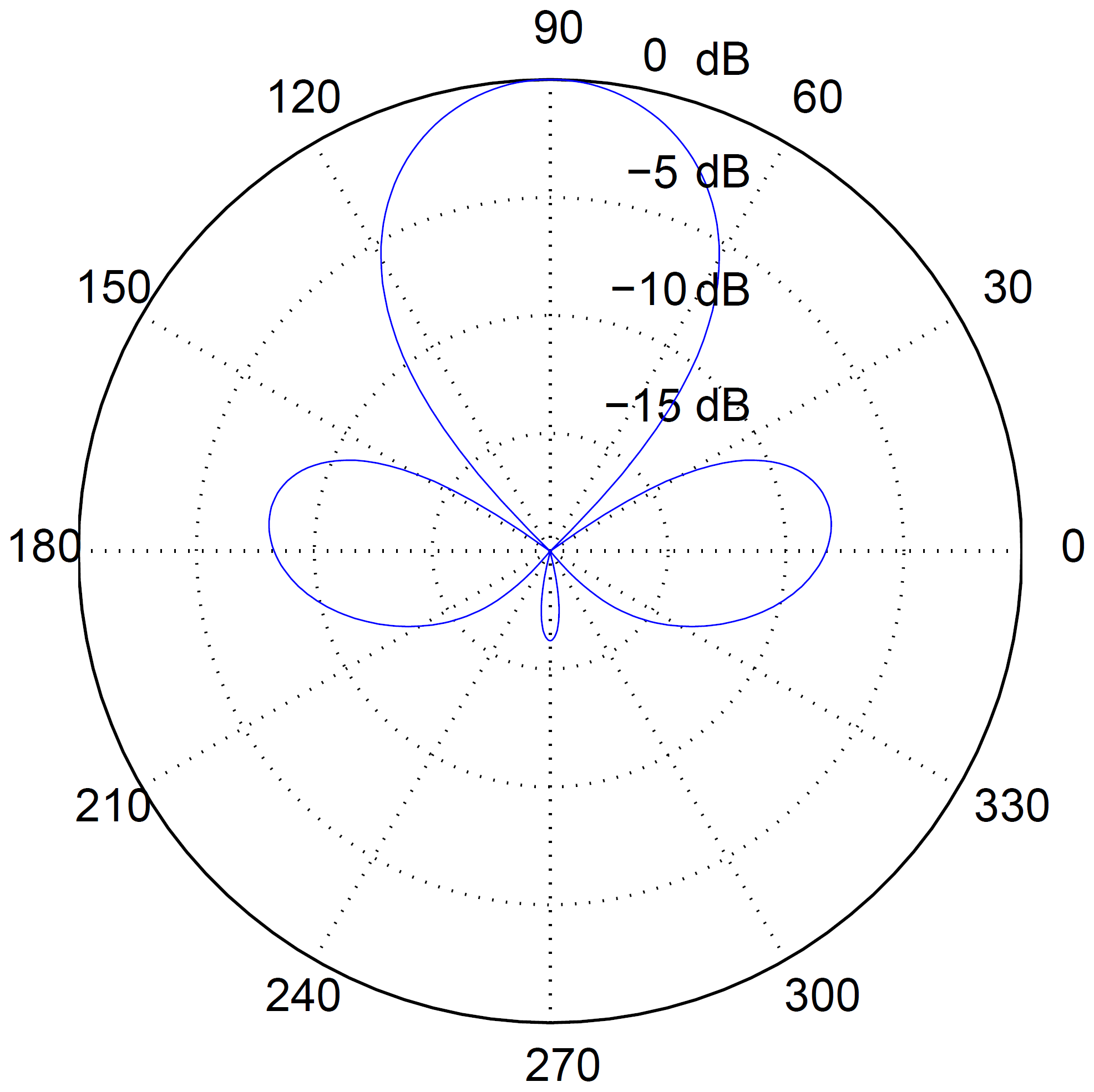} \\%
\includegraphics[width=47mm]{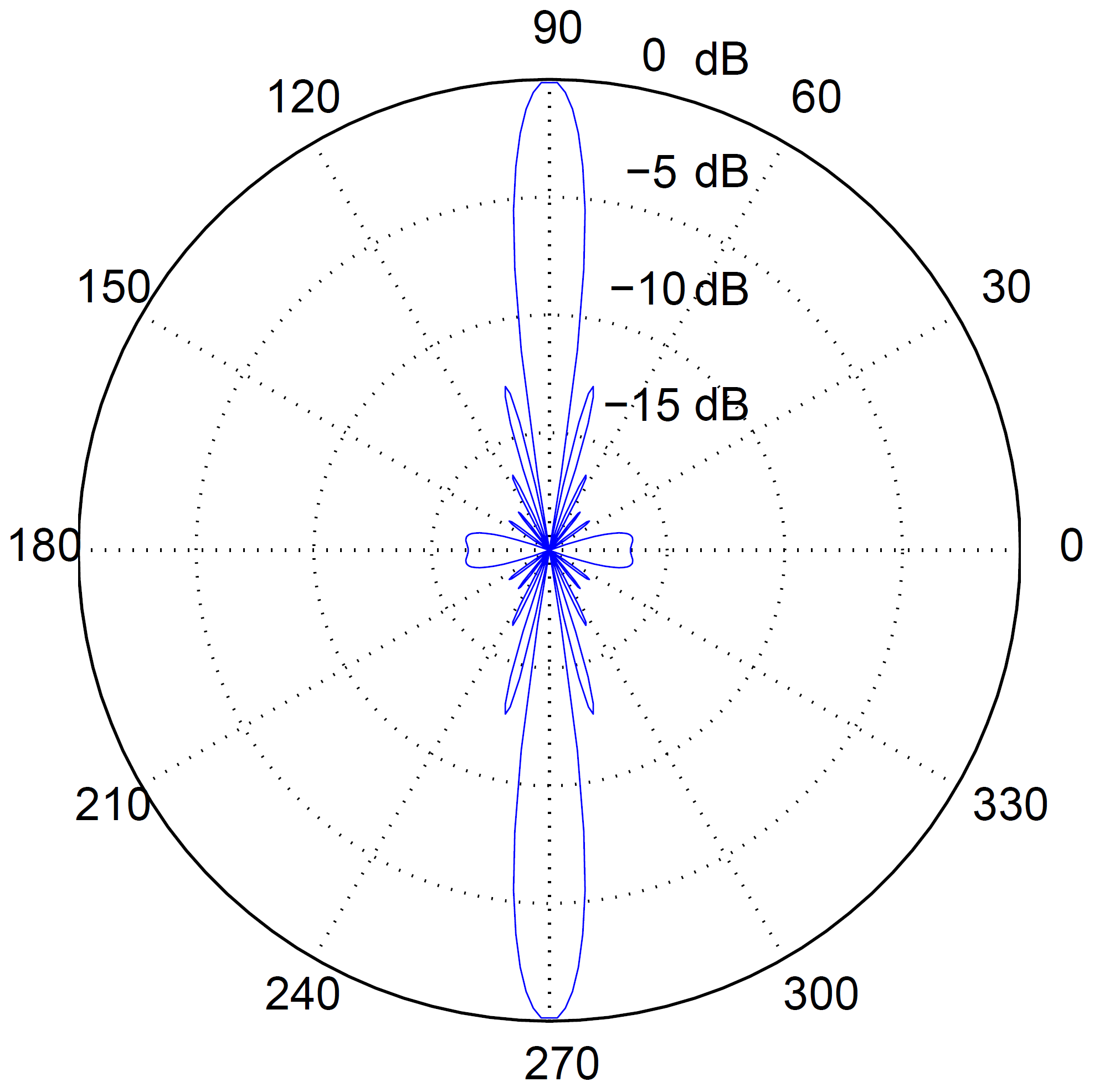}\includegraphics[width=47mm]{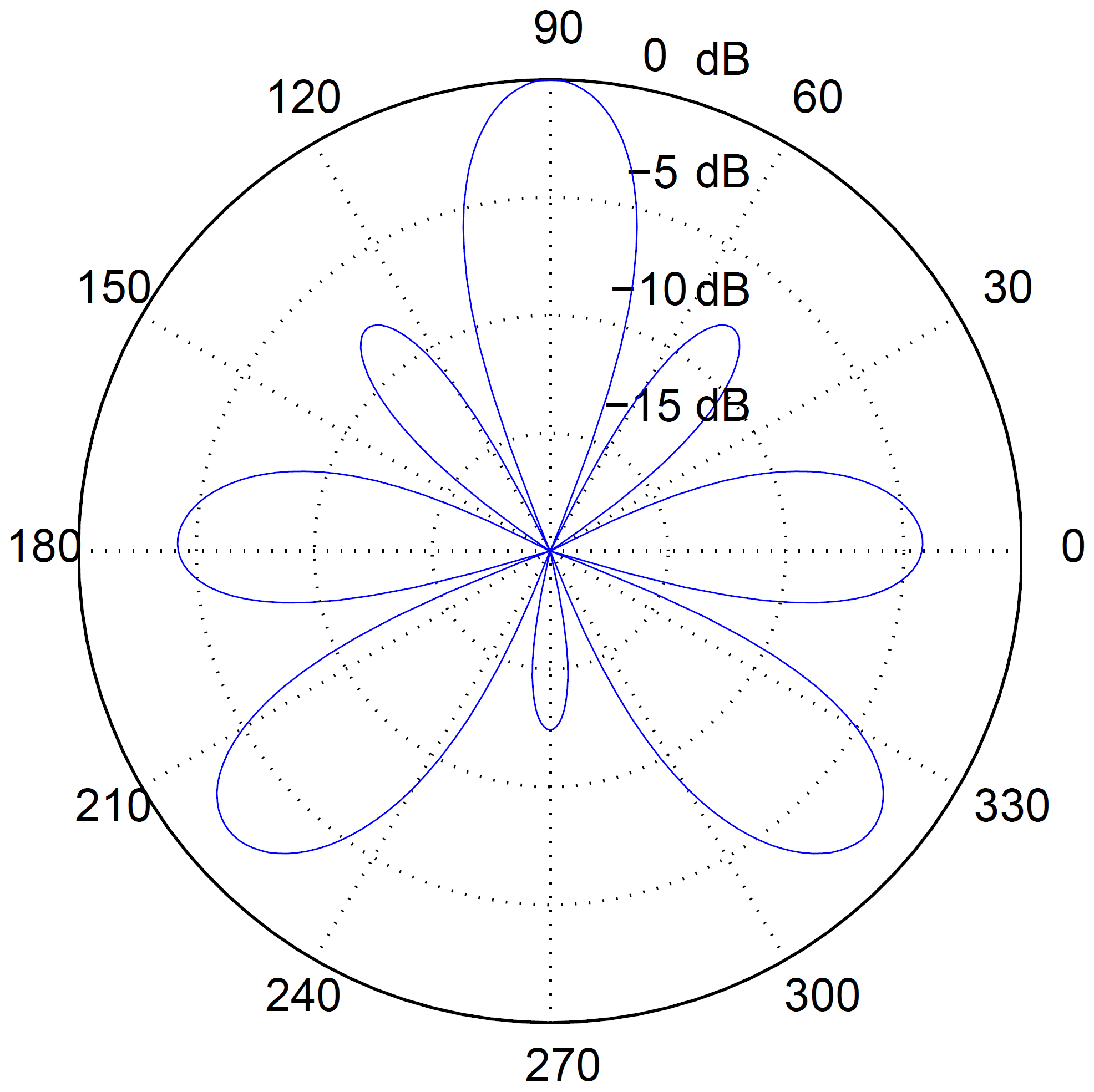}
\end{minipage}
\caption{Polar plot of the beam pattern for a linear (left) and circular (right) $8$-microphone array with $d=6$ cm and for frequencies (from top to bottom): $300$ Hz, $2000$ Hz, and $4000$ Hz.}
\label{fig_polarplot}
\end{figure}

\subsection{Linear and Circular Array Beam Pattern}

The beamformer response of a microphone array is an expedient characteristic that can be utilized to evaluate the anticipated performance of the beamformer, and it can be exploited as a gauge to indicate the capability of a beamformer to pass acoustic signals from a preferred direction and reject acoustic signals from other directions. The beamformer response $H_\phi(\omega,\theta)$ of a microphone array is also called \emph{beam pattern}. We establish here a comparison between the response of a uniform linear and circular $8$-microphone array. The beam pattern is frequency $\omega$ and DoA $\theta$ dependent, a two-dimensions
\\ \indent
The frequency response of the DaS beamformer is calculated over a significant frequency range of interest, namely, from $300$ Hz to $4000$ Hz, hence the following figures are obtained for linear and circular arrays. Figure \ref{fig_linearBP} portrays the beam pattern for $8$-microphone linear array, whereas Figure \ref{fig_circularBP} describes the beam pattern for $8$-microphone circular array. For both geometries, the distance between the microphones in the array is small ($d = 6$ cm). The simulation over $360^\circ$ in the linear array, generates two main lobes located at $90^\circ$ and $270^\circ$. While the simulation in the circular array, produces one main lobe located at $90^\circ$ and seven other lobes with negligible magnitude. Therefore, the total number of lobes in the response of the circular array is eight. In both linear and circular arrays, for a fixed beam pattern length, the beam width is decreasing much more with increasing frequency. As shown in linear array, the response at the direction of the two main lobes is unity gain in linear scale. Therefore, the beamforming is enormously enhanced for medium and high frequencies. Furthermore, there is no aliasing in the sampled acoustic signal. The beam pattern of the circular array demonstrates that the main lobe has a unity gain in linear scale at $90^\circ$ steering angle. However, the beamforming is poor at low frequencies. Other lobes such as the one at $270^\circ$ have incomparable response and they reject the acoustic signals almost in the medium frequencies.
\\ \indent
Figure \ref{fig_polarplot} brings a visual comparison between the beam pattern (in polar coordinates) for the linear and circular microphone array beamformer at the following single frequencies: $300$ Hz, $2000$ Hz, and $4000$ Hz respectively. The linear array delivers a narrow beam with low-magnitude side lobes; however, the presence of an ``ambiguity" lobe represents a severe handicap of this geometry. On the other hand, the circular array presents one single main lobe, but several side lobes or ``granting" lobes are manifestable; however those lobes are sporadic and not consistent along the desired bandwidth. The beamforming performance at low frequencies is poor in both geometries.

%\psdraft
\begin{figure}[!t]
\centering
\includegraphics[width=89mm]{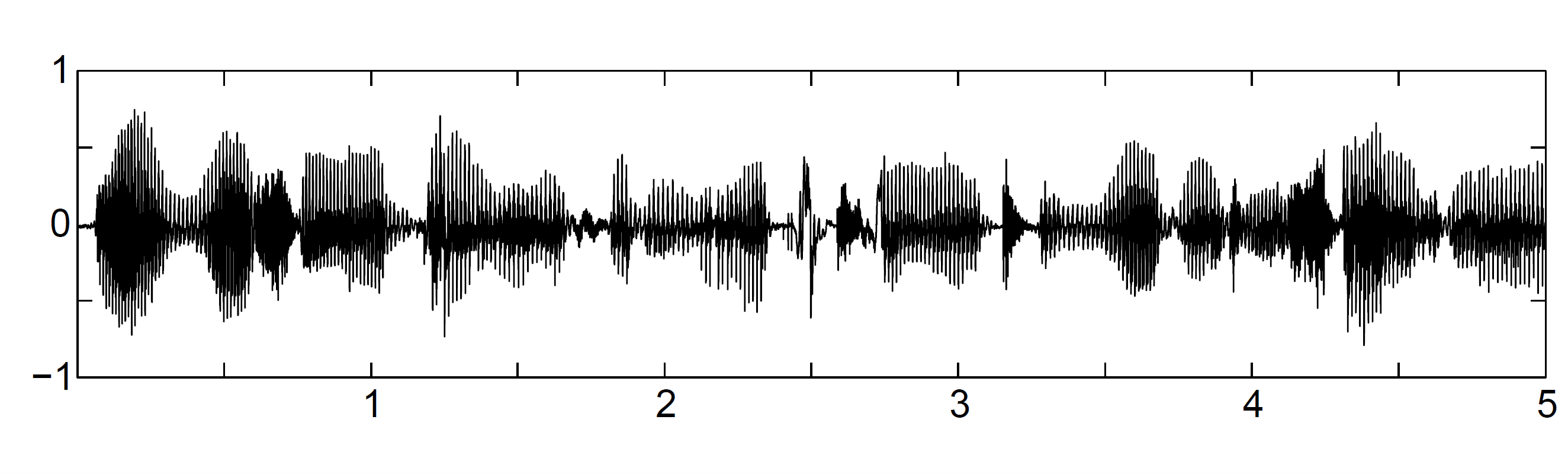} \\
\vspace{-6.5mm}\hspace{70mm} $t$ \\ \vspace{1mm}
\includegraphics[width=89mm]{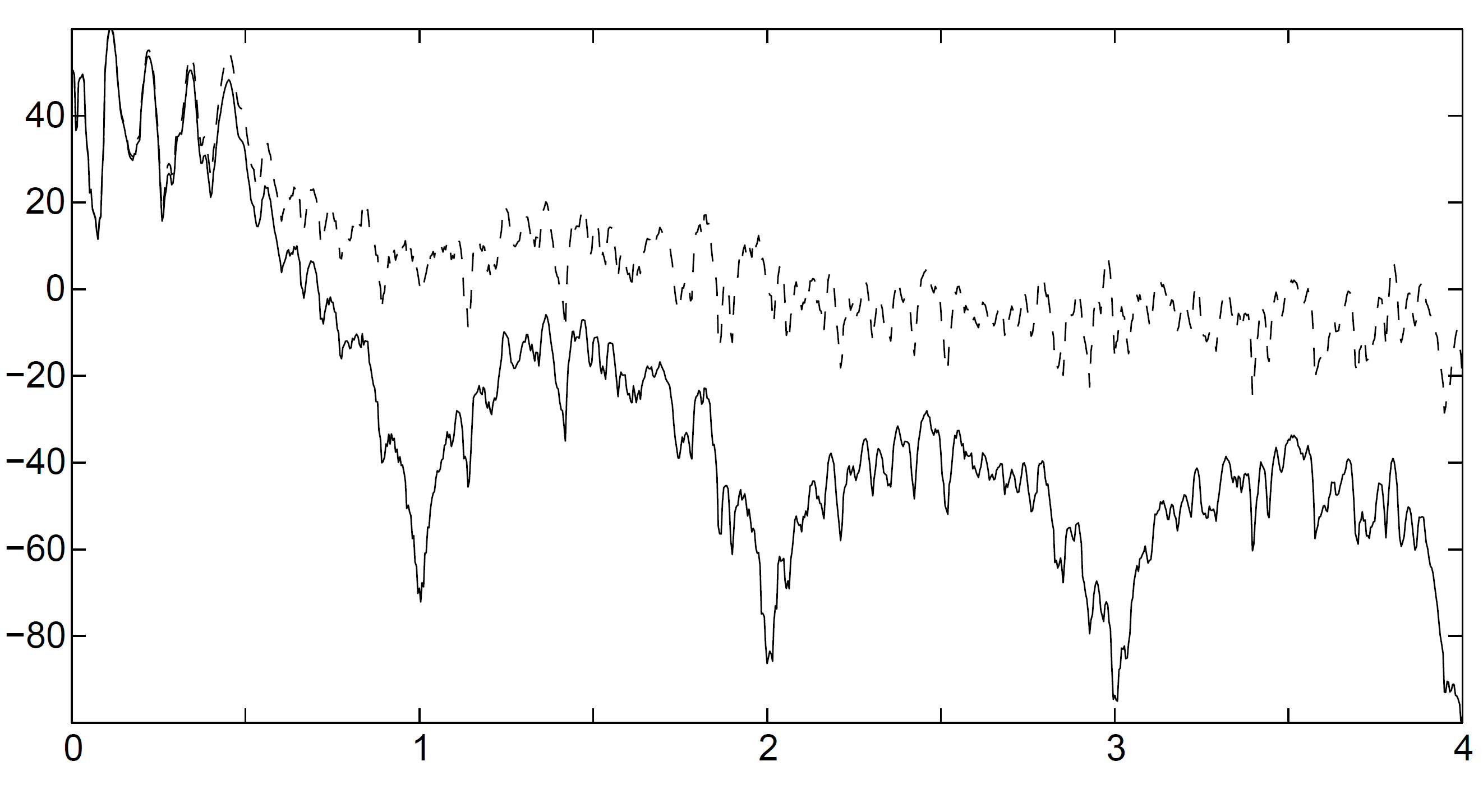} \\
\vspace{-6mm}\hspace{68mm} kHz \\ \vspace{2mm}
\includegraphics[width=89mm]{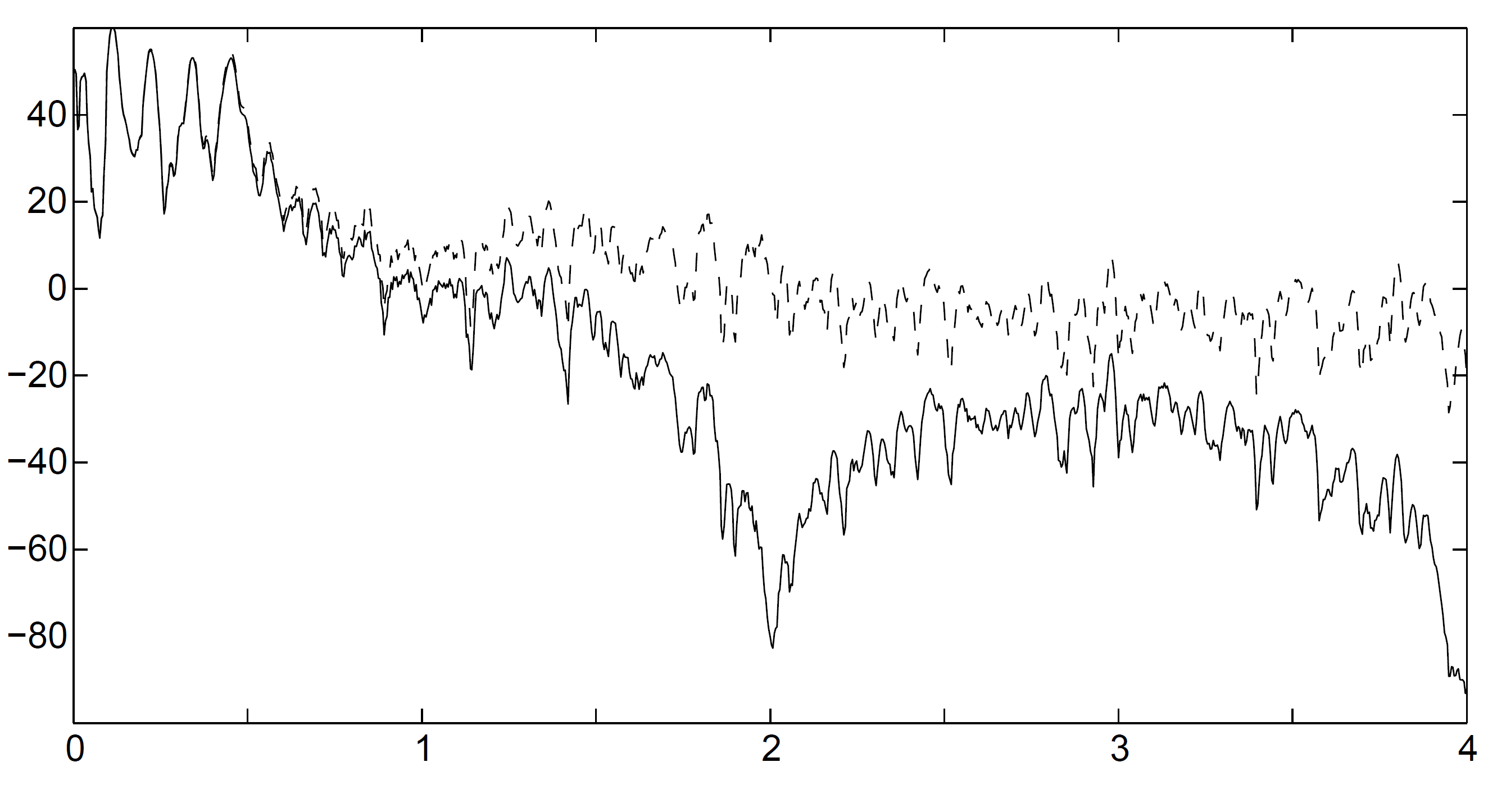} \\
\vspace{-6mm}\hspace{68mm} kHz \\ \vspace{2mm}
\caption{Spectral attenuation delivered by a microphone array with linear (middle) and circular (bottom) geometry on a real sound source (top): original spectrum (dashed) and spectrum at the beamformer output (solid). Spectrum magnitude is given in dB.}
\label{fig_speech}
\end{figure}

\subsection{Linear and Circular Array Simulation}

In order to validate the beamforming effect of the microphone array, we simulated the array mechanism as given by \eqref{average} for both geometries, linear \eqref{physdelay1}-\eqref{digdelay1} and circular \eqref{physdelay2}-\eqref{digdelay2}. The signal used in the experiments corresponds to a speech signal sampled at $16$ kHz: the sampling frequency is $4$ times the bandwidth of interest ($4$ kHz), a value required to implement the digital delays with enough accuracy. The steering angle of the array is $90^\circ$, $\phi=\pi/2$, and the sound DoA is $45^\circ$, $\theta=\pi/4$. Figure \ref{fig_speech} shows on top the original speech signal, which simulates the sound source. The two lower pictures illustrate the spectral attenuation pattern for each array geometry. The linear geometry clearly has the upper hand, with attenuation levels of $20$--$30$ dB consistently along the bandwidth. On the other hand, the circular geometry delivers good attenuation levels, reaching the $20$ dB mark, although not throughout the whole bandwidth of interest. The beamforming effect in the lowest part of the spectrum is in both cases inefficient. Nevertheless, our listening tests confirmed the efficacy of beamforming as technique for spatial discrimination of sound sources.

\section{Conclusions and Further Work}

The analysis and simulated results presented in this paper confirms the efficacy of microphone-array beamforming at attenuating sound sources coming from directions different than the predefined steering angle;  any sound source coming at this angle is captured with no distortion. This work corresponds to the analysis stage of a BSc thesis aiming at implementing in real hardware a circular microphone array. Several milestones need to be achieved, such as, mounting electret microphones and their preamplifiers in a physical circular frame, implementing the digital beamforming algorithms in a $8$-channel DSP board, and improving the signal processing algorithms. In this respect, we are exploring other beamforming techniques, such as the filter-and-sum, which can be easily integrated in the architecture under development. Finally, the circular geometry has been chosen for several reasons: it does not present ``ambiguity" lobes, it does not require much space, it is physically compact, and delivers acceptable performance within the bandwidth of speech.

\end{document}